\documentclass[dvipsnames]{article} %
\usepackage{colm2024_conference}

\usepackage{booktabs}
\usepackage{graphicx}
\usepackage{enumitem}
\usepackage{wrapfig}
\usepackage{algorithm}
\usepackage{algpseudocode}

\usepackage{graphicx}

\usepackage{microtype}
\usepackage{amsmath}
\usepackage{colortbl}
\usepackage[utf8]{inputenc}
\definecolor{lightgray}{rgb}{0.9,0.9,0.9}
\usepackage{caption}
\usepackage{subcaption}
\usepackage{setspace}
\usepackage{url}
\usepackage{multirow}
\usepackage{colortbl}
\usepackage{tabularx}
\usepackage{blindtext}
\usepackage{float}
\usepackage{pgfplots}
\pgfplotsset{compat=1.18} 
\usepackage{tikz}
\usetikzlibrary{er,positioning,bayesnet}
\usepackage{makecell}
\usepackage{tipa}
\usepackage{siunitx}
\usepackage{nicefrac}
\usepackage{tocloft}
\usepackage{listings}
\usepackage[raster,skins]{tcolorbox} %
\usepackage{xltabular}
\usepackage{adjustbox}
\usepackage{xurl}
\usepackage{rotating}
\usepackage[normalem]{ulem}
\useunder{\uline}{\ul}{}
\usepackage{hyperref}  
\usepackage{cleveref}
\usepackage{courier}  
\usepackage{booktabs}
\crefname{figure}{Figure}{Figure}
\crefname{table}{Table}{Tables}

\usepackage{fancyhdr}
\renewcommand{\headrulewidth}{0pt} 
\fancypagestyle{firstpage}{
\fancyhf{} 
\fancyhead[L]{\includegraphics[height=12mm]{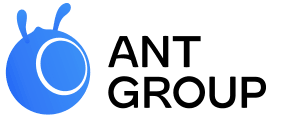}} 
\fancyhead[R]{\includegraphics[height=12mm]{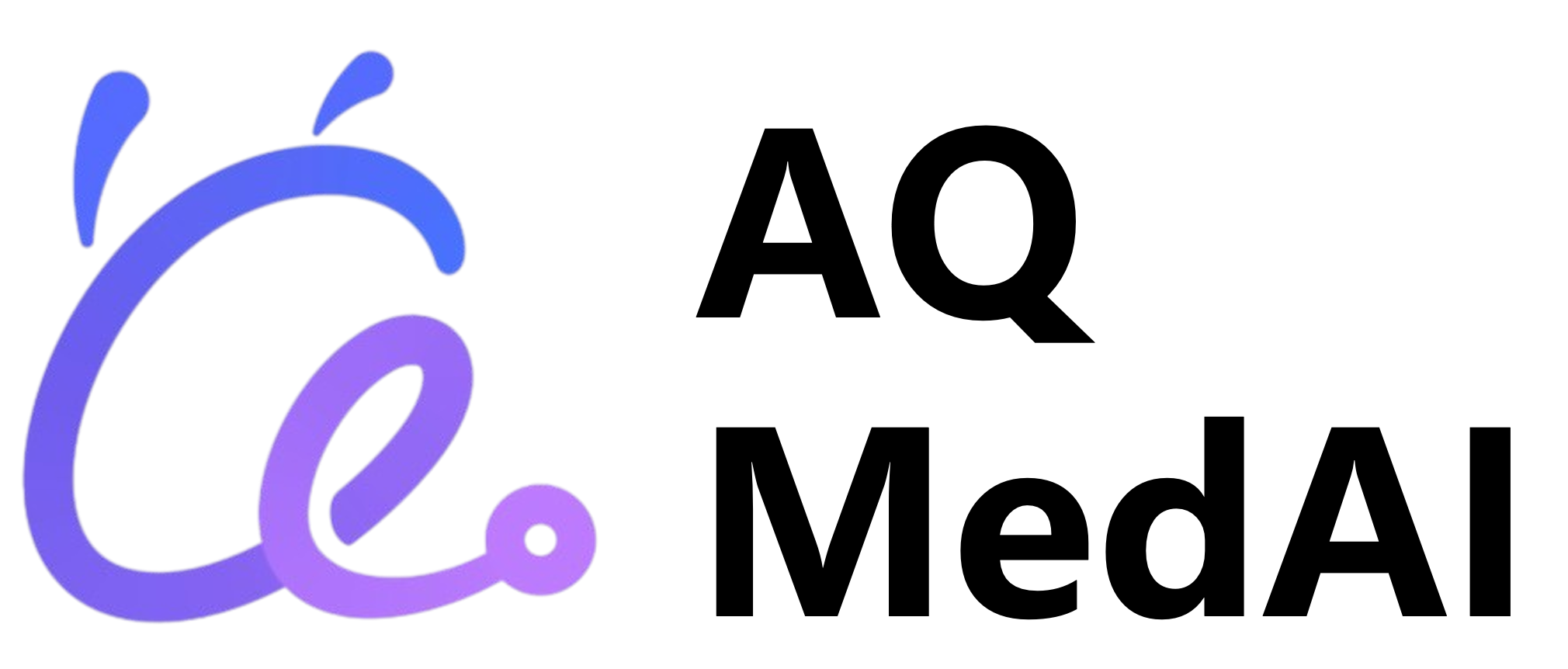}}
\fancyfoot[C]{\thepage} 
\renewcommand{\headrulewidth}{0.4pt} 
}

\newtcolorbox[auto counter]{promptbox}[2][]{
    colback=blue!5!white, 
    colframe=blue!30!black, 
    fonttitle=\bfseries, 
    title={Prompt \thetcbcounter: #2}, 
    label={#1} 
}

\usepackage{amsmath,amsfonts,bm}

\def\eqref#1{equation~\ref{#1}}

\def\1{\bm{1}}

\DeclareMathAlphabet{\mathsfit}{\encodingdefault}{\sfdefault}{m}{sl}
\SetMathAlphabet{\mathsfit}{bold}{\encodingdefault}{\sfdefault}{bx}{n}

\newcommand*\justify{%
  \fontdimen2\font=0.4em
  \fontdimen3\font=0.2em
  \fontdimen4\font=0.1em
  \fontdimen7\font=0.1em
  \hyphenchar\font=`\-
}

\renewcommand{\texttt}[1]{%
  \begingroup
  \ttfamily
  \begingroup\lccode`~=`/\lowercase{\endgroup\def~}{/\discretionary{}{}{}}%
  \begingroup\lccode`~=`[\lowercase{\endgroup\def~}{[\discretionary{}{}{}}%
  \begingroup\lccode`~=`.\lowercase{\endgroup\def~}{.\discretionary{}{}{}}%
  \catcode`/=\active\catcode`[=\active\catcode`.=\active
  \justify\scantokens{#1\noexpand}%
  \endgroup
}

\title{
DIVER: A Multi-Stage Approach for Reasoning-intensive Information Retrieval}

\author{
{\bf Duolin Sun\textsuperscript{\rm 1}, Meixiu Long\textsuperscript{\rm 12}\thanks{Work done during the internship at Ant Group.}, Dan Yang\textsuperscript{\rm 1}\thanks{Corresponding author}, Junjie Wang\textsuperscript{\rm 1},Yecheng Luo\textsuperscript{\rm 1}, Yue Shen\textsuperscript{\rm 1}, Jian Wang\textsuperscript{\rm 1}, Hualei Zhou\textsuperscript{\rm 1}, Chunxiao Guo\textsuperscript{\rm 1}, Peng Wei\textsuperscript{\rm 1}, Jiahai Wang\textsuperscript{\rm 2}, Jinjie Gu\textsuperscript{\rm 1}} \\
\textsuperscript{1}Ant Group, Hangzhou, China
\textsuperscript{2}Sun Yat-sen University, Guangzhou, China
 \\ 
\texttt{}
\texttt{\{longmx7\}@mail2.sysu.edu.cn, \{sunduolin.sdl, luoyin.yd\}@antgroup.com}
}

\begin{document}

\maketitle

\begin{tcolorbox}[colback=blue!5!white,  
                  colframe=blue!20!white,          
                  boxrule=2pt,            
                  arc=3mm]                

\begin{abstract}
\vspace{5pt}
Retrieval-augmented generation has achieved strong performance on knowledge-intensive tasks where query-document relevance can be identified through direct lexical or semantic matches. However, many real-world queries involve abstract reasoning, analogical thinking, or multi-step inference, which existing retrievers often struggle to capture.
To address this challenge, we present \textbf{DIVER}, a retrieval pipeline designed for reasoning-intensive information retrieval. It consists of four components. 
The document preprocessing stage enhances readability and preserves content by cleaning noisy texts and segmenting long documents. The query expansion stage leverages large language models to iteratively refine user queries with explicit reasoning and evidence from retrieved documents. The retrieval stage employs a model fine-tuned on synthetic data spanning medical and mathematical domains, along with hard negatives, enabling effective handling of reasoning-intensive queries. Finally, the reranking stage combines pointwise and listwise strategies to produce both fine-grained and globally consistent rankings.
\textbf{On the BRIGHT benchmark, DIVER achieves state-of-the-art nDCG@10 scores of 46.8 overall and 31.9 on original queries, consistently outperforming competitive reasoning-aware models.} These results demonstrate the effectiveness of reasoning-aware retrieval strategies in complex real-world tasks.


\vspace{5pt}
\textbf{Code}: \url{https://github.com/AQ-MedAI/Diver}

\textbf{Model}: \url{https://huggingface.co/AQ-MedAI/Diver-Retriever-4B-1020}, \url{https://huggingface.co/AQ-MedAI/Diver-GroupRank-32B}


\end{abstract}
\end{tcolorbox}

\begin{figure}[H]
    \centering
    \includegraphics[width=0.9\textwidth]{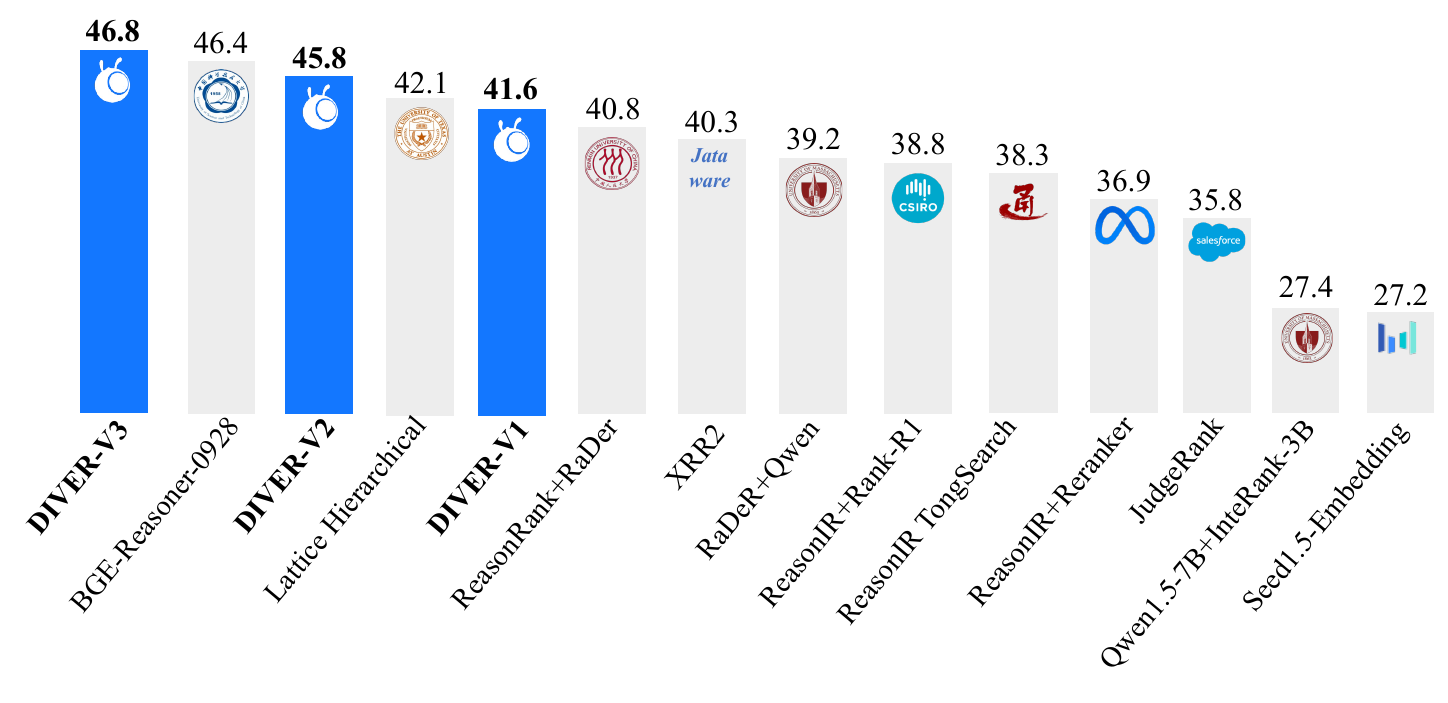}
    \caption{DIVER achieves state-of-the-art performance on BRIGHT benchmark.}
    \label{fig:bright_bench}
\end{figure}


\newpage

\section{Introduction}

\label{sec:intro}
Retrieval-augmented generation (RAG) has been widely applied in knowledge-intensive tasks such as factual question answering~\citep{factQA_survey,metrag_2024,jiao2025hirag,tan2025prgb,gan2025polyrag}. In these tasks, relevant documents that directly address the query can often be retrieved through simple lexical or semantic matching.
However, relevance is not always established through surface-level similarity. In some cases, it arises from deeper and more abstract relationships, such as shared reasoning patterns or analogous conceptual structures. Real-world queries are often complex and require more than superficial similarity to identify relevant information. For example, an economist may seek a case study that illustrates the same economic principle as another, or a programmer may aim to resolve an error by finding documentation that explains the corresponding syntax. In such cases, effective retrieval demands reasoning beyond lexical or embedding-based similarity.

To better evaluate retrieval models under such challenging conditions, the BRIGHT benchmark~\citep{25ICLR_BRIGHT} was introduced and has gained significant attention. It consists of 1,384 real-world queries collected from domains including economics, psychology, mathematics, and programming, all sourced from authentic human-generated data.
Unlike earlier rag benchmarks such as BEIR~\citep{BEIR_benchmark}, PRGB~\cite{tan2025prgb} MTEB~\citep{MTEB_benchmark}, which mainly target fact-based queries commonly derived from search engines and resolved through direct keyword or embedding-based matching, BRIGHT focuses on reasoning-intensive retrieval. The relevance between queries and documents in BRIGHT often involves implicit connections that require deliberate multi-step reasoning.

In this work, we propose \textbf{DIVER} (\underline{D}eep reason\underline{i}ng retrie\underline{v}al and r\underline{er}anking), a retrieval pipeline composed of four main components: document preprocessing, document-interactive query expansion, reasoning-enhanced retrieval, and hybrid pointwise-listwise reranking.
The first stage addresses document quality. Many of the provided documents suffer from poor readability, with issues such as excessive blank lines and truncated sentences. To improve input quality, the documents are cleaned. Additionally, a subset of documents exceeds the 16k token limit of the encoder. Rather than truncating them, which risks information loss, the documents are rechunked into segments of up to 4k tokens. This preprocessed version is referred to as \textbf{DIVER-DChunk}.
The second component, \textbf{DIVER-QExpand}, utilizes an explicit reasoning chain generated by a large language model (LLM) to iteratively expand and refine the user query. Through multiple rounds of interaction with retrieved documents, the query is dynamically updated based on newly retrieved evidence, enabling diverse and context-aware interpretations of the original query.
The third component focuses on reasoning-intensive retrieval. Existing retrievers, typically trained on datasets with short, fact-based queries, struggle with tasks requiring complex reasoning, as seen in benchmarks like BRIGHT. To address this, we construct synthetic training data by combining medical, coding, and mathematical domains, along with hard negative documents. This data is used to fine-tune Qwen3-Embedding-4B~\citep{qwen3emb} under a contrastive learning objective, resulting in a retriever specifically adapted to complex reasoning, termed as \textbf{DIVER-Retriever}. To further enhance retrieval quality, relevance scores from this retriever are interpolated with BM25 scores, capturing both deep reasoning and surface-level similarities.
The final component, \textbf{DIVER-Rerank}, performs pointwise reranking of retrieved documents. An off-the-shelf LLM assigns each document an integer helpfulness score (e.g., from 0 to 10). As LLM-generated scores often produce ties, they are interpolated with the retrieval scores to yield more fine-grained and discriminative rankings. In addition, DIVER-Rerank integrates the pointwise reranker with a listwise reranker, capturing both local relevance and a holistic ranking perspective.


Comprehensive experiments are conducted on the BRIGHT benchmark. Experimental results show that DIVER achieved an improved nDCG@10 of 46.8, surpassing the BGE-Reasoner and establishing a new SOTA.  When using only the original queries, DIVER obtains an nDCG@10 of 31.9, outperforming strong reasoning-intensive baselines including Seed1.5-Embedding\footnote{\url{https://seed1-5-embedding.github.io/}}, ReasonIR~\citep{25_ReasonIR}, and RaDeR~\citep{25_RaDeR}. These results validate the effectiveness of both the retrieval strategy and the synthesized training data.

\section{DIVER Model}
\begin{figure}[tbp]
    \centering
    \includegraphics[width=0.9\linewidth]{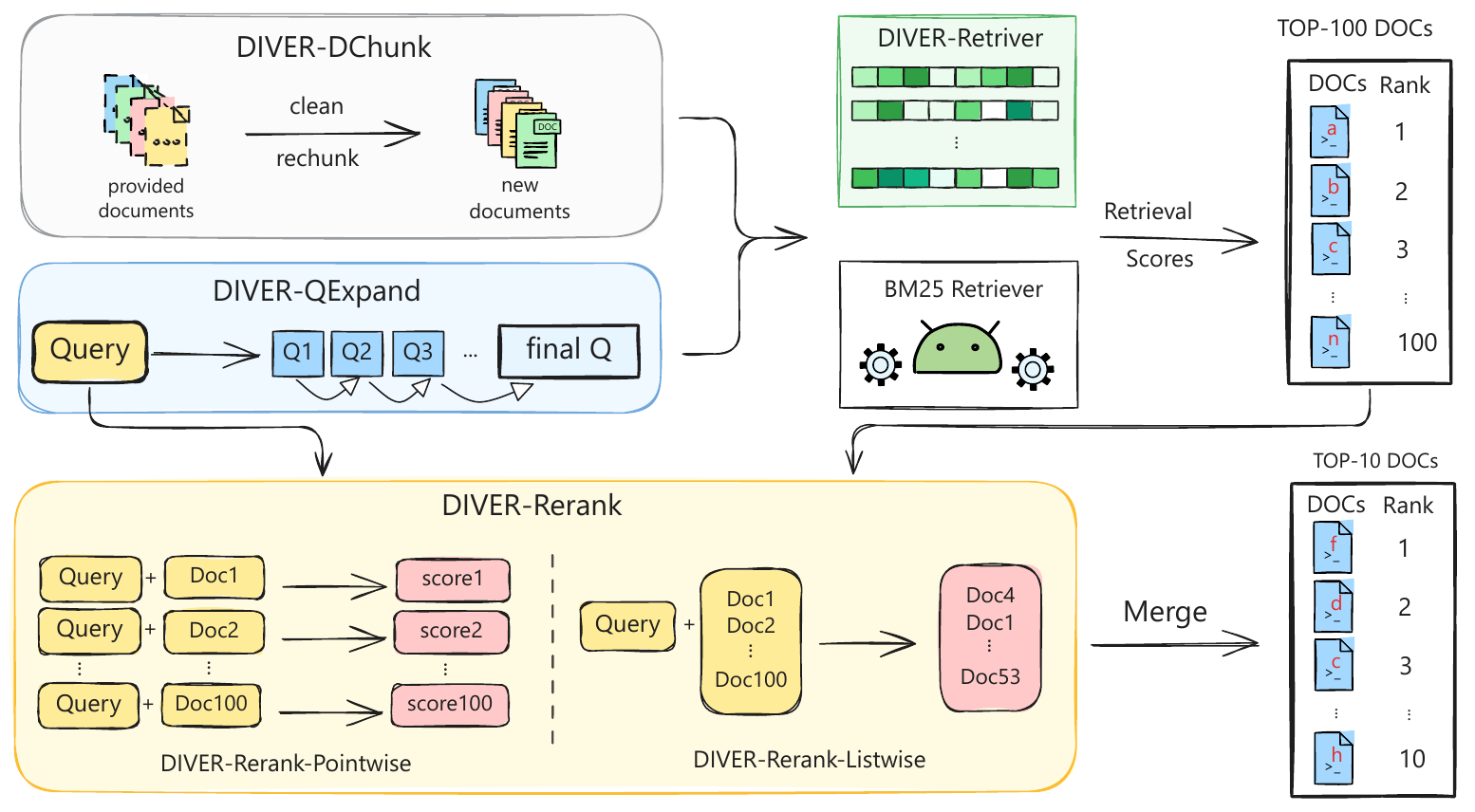}
    \caption{Overview of DIVER pipeline. The DIVER pipeline begins with document cleaning and semantic-based rechunking to improve textual coherence. User queries are then iteratively expanded using the DIVER-QExpand module to enhance their expressiveness. Document relevance is scored by both the DIVER Retriever and BM25 Retriever. The top-100 candidates are reranked using DIVER-Rerank, which assigns LLM-based helpfulness scores. Final rankings are obtained by interpolating pointwise and listwise reranking scores to improve overall precision. To ensure fair comparison with other baselines, DIVER-DChunk is excluded from the main experiments and only evaluated separately in \cref{sec:ablation_diver_dchunk}.}
    \label{fig:overview_diver}
\end{figure}

This section details the DIVER framework, which comprises document processing, a first-stage query expansion and retrieval module, and a second-stage reranking model. An overview of the pipeline is illustrated in \cref{fig:overview_diver}.

\subsection{Document Processing}
The BRIGHT benchmark covers three high-level domains with StackExchange, Coding, and Theorem-based content. We observe that significant data quality issues are concentrated in seven StackExchange-derived subdomains. These issues, characteristic of web-scraped content, include truncated sentences, excessive blank lines, and other structural inconsistencies that hinder semantic coherence. For retrievers, this is akin to reading a disordered book cluttered with boilerplate, disrupting narrative flow and making reasoning more difficult.

To address these issues and enhance document readability, we implemented a rule-based cleaning process. This process involved removing unnecessary blank lines, merging incomplete sentences, and eliminating redundant spaces, followed by reassembling paragraphs to restore logical flow and structural coherence. 

Additionally, some documents exceeded the 16k token limit of certain encoders, and long text chunks often included semantically irrelevant information, further impairing retrieval accuracy and downstream reasoning.
To handle lengthy documents, we employed the \texttt{Chonkie}\footnote{\url{https://github.com/chonkie-inc/chonkie}} library to perform semantic-aware chunking. Using the Qwen3-Embedding-0.6B~\citep{qwen3emb} model with a similarity threshold of 0.5, the text was divided into smaller chunks of up to 4096 tokens, with a minimum size of one sentence per chunk. To ensure semantic continuity across chunks, we applied an overlap refinement strategy that retained 20\% overlapping content using a character-based suffix method. This approach achieves a balance between maintaining semantic integrity and capturing sufficient context for downstream tasks.
To ensure consistency with the BRIGHT benchmark ground truth, we retained the original document IDs for each chunk instead of reassigning new IDs. As a result, some document IDs may correspond to multiple chunks. In subsequent computations of similarity between queries and documents, the relevance score is assigned by taking the maximum similarity among all chunks that share the same ID. This ensures alignment with the benchmark while enabling fine-grained document processing.

\begin{table}[h]
\centering
\caption{Prompts used in DIVER-QExpand for query expansion. Braces \{\} denote placeholders.}
\small
\begin{tabular}{@{}p{0.22\linewidth}p{0.73\linewidth}@{}}
\toprule
\textbf{Prompt Stage} & \textbf{LLM Instruction} \\
\midrule
\textbf{First Round} & Given a query and the provided passages (most of which may be incorrect or irrelevant), identify helpful information from the passages and use it to write a correct answering passage. Use your own knowledge, not just the example passages! \\[0.5em]
& Query: \texttt{\{query\}}\\
& Possible helpful passages: \texttt{\{top-k retrieved documents\}} \\
\midrule
\textbf{Subsequent Rounds} & Given a query, the provided passages (most of which may be incorrect or irrelevant), and the previous round's answer, identify helpful information from the passages and refine the prior answer. 
Ensure the output directly addresses the original query. Use your own knowledge, not just the example passages!  \\[0.5em]
& Query: \texttt{\{query\}}\\
& Possible helpful passages: \texttt{\{top-k retrieved documents\}}\\
& Prior generated answer: \texttt{\{last-round expansion\}}\\
\bottomrule
\end{tabular}
\label{tab:query_expand_prompt}
\end{table}

\subsection{Query Expansion}
Query expansion \citep{wang-etal-2024-learning-plan} is a widely used technique in information retrieval, aiming to improve performance by enriching initial queries with contextually relevant terms. ThinkQE~\citep{25_ThinkQE} is a query expansion framework that integrates LLM-based reasoning with iterative corpus interaction. The process consists of multiple rounds of query expansion and document retrieval. In each round, the LLM generates an expanded query based on the newly retrieved documents, which then guides the retrieval of more relevant documents in the next round, forming a feedback loop for progressive refinement.

Based on ThinkQE~\citep{25_ThinkQE}, we retain its iterative design in DIVER-QExpand but make two practical modifications. First, we replace the BM25 retriever with a dense retriever trained for reasoning-intensive task (see Section~2.3). Its semantic similarity scores yield passages that are more relevant to the query's intent. Second, instead of concatenating all intermediate query expansions, which can often exceed 2000 tokens, we simplify the process by retaining only the original query and the final-round expansion. This approach reduces length of expanded query while maintaining critical information for effective retrieval. 

We use the QWEN-R1-Distill-14B model~\citep{25_Deepseek_R1} with a temperature of 0.7 to generate expansions with controlled variability. The prompt instructions vary depending on whether it is the first or a subsequent round, as shown in Table~\ref{tab:query_expand_prompt}.
DIVER-QExpand performs two rounds of retrieval and expansion, selecting the top-5 documents in each. To ensure diversity, previously retrieved documents are excluded in later rounds. Each document is truncated to 512 tokens before being fed into the LLM. These adjustments balance computational efficiency with retrieval performance.

\begin{figure}[tbp]
    \centering
    \includegraphics[width=\textwidth]{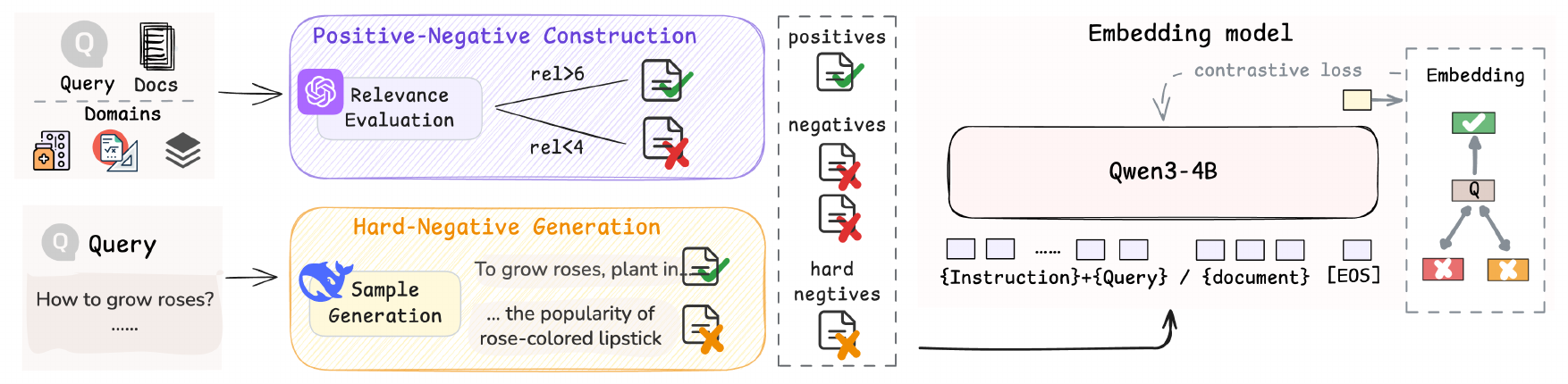}
    \caption{The training process of DIVER-Retriever. }
    \label{fig:diver_retriever}
\end{figure}

\subsection{Reasoning-intensive Retriever}
Most existing retrievers are trained on datasets consisting of short factual queries, where the relationship between queries and relevant documents is straightforward. This reliance on such datasets limits their performance on reasoning-intensive retrieval tasks, which require deeper understanding and more complex reasoning. For instance, the leading SFR-EmbeddingMistral~\citep{SFR-Embedding-Mistral} achieves a score of 59.0 on the MTEB~\citep{MTEB_benchmark} retrieval subset (BEIR~\citep{BEIR_benchmark}) but performs significantly worse on reasoning-intensive benchmarks such as BRIGHT, where the nDCG@10 score drops to 18.3.

To address these performance gaps, ReasonIR-8B~\citep{25_ReasonIR} was introduced as the first bi-encoder retriever tailored for reasoning-intensive tasks. This model employs a synthetic data generation pipeline that produces queries of varying lengths and reasoning complexity for each document while also generating challenging hard negatives—documents that appear superficially relevant but lack actual relevance. Another approach, RaDeR~\citep{25_RaDeR}, proposes a family of dense retrievers trained using data derived from mathematical problem-solving processes generated by LLM. RaDeR enhances training through retrieval-augmented reasoning trajectories and self-reflective relevance evaluations, which help generate diverse and hard negative examples. Although primarily trained on mathematical reasoning data, RaDeR models generalize well to other reasoning tasks in the BRIGHT benchmark, showing particularly strong performance on the Math and Coding subsets.

Both REASONIR-8B and RaDeR highlight a critical challenge in developing reasoning-based retrievers: creating high-quality datasets that encompass diverse query formats, lengths, and reasoning complexities. Therefore, we generate training data that incorporates synthetic hard negatives from the medical, general and mathematical domains, thereby increasing both diversity and difficulty. Specifically, 60,000 medical examples and 20,000 general examples are collected from real-world cases, with an additional 20,000 hard negatives generated using the strategy described in \cref{sec:appendix}. 
For the mathematical domain, we include problem-solving trajectories\footnote{\url{https://huggingface.co/datasets/Raderspace/MATH_NuminaMath_allquerytypes}}, comprising about 120,000 examples with hard negatives, to enhance performance on math-related reasoning tasks. The training process of DIVER-Retriever is illustrated in \cref{fig:diver_retriever}, and further details on the construction of positive and negative samples are presented in \cref{sec:appendix}.

Our dense retriever adopts Qwen3-Embedding-4B~\citep{qwen3emb} as its backbone, a powerful pretrained language model that excels in various language tasks. To obtain a vector representation for a given text, we feed the full input sequence into Qwen3-Embedding-4B and extract the hidden state of the end-of-sequence (EOS) token from the model’s last layer. The hidden state of [EOS] token is theoretically expected to capture the complete contextual and semantic information of the sequence, as it reflects the model’s final internal state after processing all input tokens. This approach is both common and effective in contrastive-learning frameworks as well as other text-embedding tasks.

The retriever is fine-tuned using the InfoNCE loss~\citep{infonce}, commonly used in contrastive learning. It minimizes the distance between semantically similar pairs while maximizing the distance from dissimilar pairs. Given an input sequence of tokens $t = {t_1, t_2, \ldots, t_k}$ with an appended EOS token, the decoder-only language model encodes it into an embedding $E_{t}$ using the hidden state of the EOS token, for either a query $q$ or a document $d$:
\begin{equation}
E_{t} = \text{Decoder}(t_{1}t_{2} \cdots t_{k}\langle \text{eos} \rangle)[-1],
\end{equation}
where $\text{Decoder}(\cdot)$ denotes the backbone model. Query and document embeddings are compared via cosine similarity: $s(q, d) = \cos(E_q, E_d)$. 
The InfoNCE loss is computed as:
\begin{equation}
\mathcal{L}(q, d^+, D^{-}) = -\log\frac{\exp(s(q, d^+))}{\exp(s(q, d^+)) + \sum_{d^- \in D^{-}}\exp(s(q, d^-))},
\end{equation}
where $d^+$ is a positive document and $D^-$ is a curated set of hard negatives. Using curated hard negatives rather than all irrelevant documents improves training efficiency without sacrificing retrieval quality.

The final relevance score between a query and a document, \(S_{retriever}\), is calculated as a weighted sum: \(S_{retriever} = 0.5 \cdot S_{DIVER-Retriever} + 0.5 \cdot S_{BM25}\). To ensure consistent scaling, both \(S_{DIVER-Retriever}\) and \(S_{BM25}\) are normalized to the range of 0 to 1. This approach effectively combines the semantic strengths of dense and sparse retrievers, improving retrieval performance as shown in \cref{sec:perf_different_retriever}.

\subsection{Reranking}
The retrieve-then-rerank paradigm is widely used to enhance retrieval performance, especially in reasoning-intensive tasks where LLM rerankers have demonstrated significant effectiveness~\citep{25ICLR_BRIGHT}. Reranking approaches typically fall into four categories: pointwise~\citep{24_pointwise1}, pairwise~\citep{24_pairwise1,25_pairwise2}, setwise~\citep{25_rankr1}, and listwise methods~\citep{24_listwise1,25_Rank_k,25_rearank}. For a more comprehensive discussion, we refer readers to \citep{25_RefRank}.

Our DIVER-Rerank integrates pointwise and listwise paradigms to capture both local and global perspectives. The pointwise module (DIVER-Rerank-Pointwise) follows the idea of ReasonIR~\citep{25_ReasonIR}, which uses an LLM (i.e., Qwen-2.5-32B-Instruct~\citep{qwen2.5}) to assign helpfulness scores to individual documents. Instead of the 0–5 scale in ReasonIR, we adopt a finer 0–10 scale to evaluate query–document relevance, normalize the scores to $[0,1]$, and compute the final pointwise score as $0.6 \cdot S_{reranker} + 0.4 \cdot S_{retriever}$ to balance retriever and reranker contributions.

To complement this local evaluation, the listwise module (DIVER-Rerank-Listwise) employs an LLM (e.g. Deepseek-R1-0528) to directly rank the top-100 candidate documents of the query, offering a global view of document relevance. The final reranking result integrates both modules, leveraging the fine-grained local scoring of pointwise rerankers and the holistic ranking ability of listwise rerankers.

\section{Experiments}
The DIVER approach is evaluated on the BRIGHT benchmark against competitive baselines. Following prior work \citep{25ICLR_BRIGHT}, the normalized Discounted Cumulative Gain at top-10 (nDCG@10) is used as the evaluation metric. Results are reported for all BRIGHT datasets: Biology (Bio.), Earth Science (Earth.), Economics (Econ.), Psychology (Psy.), Robotics (Rob.), Stack Overflow (Stack.), Sustainable Living (Sus.), LeetCode (Leet.), Pony, AoPS, and TheoremQA with question retrieval (TheoQ.) and theorem retrieval (TheoT.). Avg. denotes the average score across the 12 datasets. Baseline results are taken from their original papers or reproduced using our own implementations.

\subsection{Overall Performance}
We compare our model with competitive approaches listed on the BRIGHT leaderboard\footnote{\url{https://brightbenchmark.github.io/}}, where the top-8 methods all adopt reranking strategies to achieve superior performance. Following the standard setup in \citep{25ICLR_BRIGHT,25_ReasonIR}, reranking is applied to the top-100 documents retrieved in the first stage. This section compares DIVER with 7 baselines:
(1) \textbf{Rank-R1-14B}~\citep{25_rankr1}, a LLM-based setwise reranker designed for reasoning-intensive ranking tasks, applied to BM25 results on original queries.
(2) \textbf{Qwen1.5-7B with InteRank-3B}~\citep{25_interank} uses small-size models for reasoning-based document reranking.
(3) \textbf{GPT4 with Rank1-32B}~\citep{25_rank1}, which combines GPT-4 Reason-query with a reasoning reranker that ``think'' before making relevance judgments.
(4) \textbf{ReasonIR with QwenRerank}~\citep{25_ReasonIR}, the baseline introduced in the ReasonIR paper, which scores documents from 0 to 5.
(5) \textbf{ReasonIR with Rank-R1-32B}\footnote{\url{https://huggingface.co/ielabgroup/Rank-R1-32B-v0.2}}, ranked 5th, replaces the original pointwise reranker with with a listwise one.
(6) \textbf{RaDeR with QwenRerank}\footnote{\url{https://github.com/Debrup-61/RaDeR}}, ranked 4th, adopts the same reranker as ReasonIR.
(7) \textbf{XRR2}\footnote{\url{https://github.com/jataware/XRR2}}, currently ranked 3rd, uses prompt-based query expansion and performs multiple reranking rounds by LLM.
(8) \textbf{ReasonRank}~\citep{liu2025reasonrank}, currently ranked 2nd, proposes a RL–based listwise reranker applied to the retrieval results of RaDeR.
(9) \textbf{BGE-Reasoner}~\footnote{\url{https://github.com/FlagOpen/FlagEmbedding/tree/master/research/BGE_Reasoner}}currently ranked 1st, generates multiple rewritten queries for retrieval and ensembles reranking results from different model sizes to produce the final ranking.

\cref{tab:query_retriever_comparison} demonstrates that DIVER(v1) achieves an nDCG@10 of \textbf{41.6} on Aug 12, 2025, surpassing XRR2 by \textbf{+1.3 points} and setting a new SOTA on the BRIGHT leaderboard at that time. XRR2 relies on GPT-4o for query expansion and gemini-2.5-flash-preview-04-17 for reranking, performing multiple reranking passes with score averaging. In contrast, DIVER achieves higher performance with significantly lower computational cost. DIVER also outperforms ReasonIR and RaDeR by a substantial margin, demonstrating the effectiveness of its query expansion and retrieval pipeline.
Subsequently, on Aug 26, 2025, DIVER(v2) reaches an nDCG@10 of \textbf{45.8}, surpassing BGE-Reasoner by \textbf{+0.8 points} and establishing a new SOTA. Our improvements attribute to enhanced query expansion and the combined pointwise and listwise reranker.

\begin{table}[tbp]
\centering
\caption{Performance comparisons with competitive baselines on the BRIGHT leaderboard. The best result for each dataset is highlighted in \textbf{bold}. DIVER(v1) denotes the initial version with DIVER-QExpand, DIVER-Retriever, and DIVER-Rerank-Pointwise. DIVER(v2) includes advanced query expansion and combined pointwise and listwise DIVER-Rerank. DIVER(v3) uses BGE-Reasoner-Embed-0928~\citep{25_BGE_Reasoner_Embed} for retrieval and our proposed groupwise reranking~\citep{25_grouprank}, achieving the latest state-of-the-art performance.}
\resizebox{\textwidth}{!}{
\begin{tabular}{l|c|ccccccc|cc|ccc}
\toprule &
& \multicolumn{7}{c|}{\textbf{StackExchange}} 
& \multicolumn{2}{c|}{\textbf{Coding}} 
& \multicolumn{3}{c}{\textbf{Theorem-based}} \\
\textbf{Method} & \textbf{Avg.} 
& \textbf{Bio.} & \textbf{Earth.} & \textbf{Econ.} & \textbf{Psy.} & \textbf{Rob.} & \textbf{Stack.} & \textbf{Sus.}
& \textbf{Leet.} & \textbf{Pony} 
& \textbf{AoPS} & \textbf{TheoQ.} & \textbf{TheoT.} \\
\midrule
Rank-R1-14B & 20.5 & 31.2 & 38.5 & 21.2 & 26.4 & 22.6 & 18.9 & 27.5 & 9.2 & 20.2 & 9.7 & 11.9 & 9.2 \\ \
Qwen1.5-7B with InteRank-3B & 27.4 & 51.2 & 51.4 & 22.4 & 31.9 & 17.3 & 26.6 & 22.4 & 24.5 & 23.1 & 13.5 & 19.3 & 25.5 \\ \
GPT4 with Rank1-32B & 29.4 & 49.7 & 35.8 & 22.0 & 37.5 & 22.5 & 21.7 & 35.0 & 18.8 & 32.5 & 10.8 & 22.9 & 43.7 \\ \
ReasonIR with QwenRerank & 36.9 & 58.2 & 53.2 & 32.0 & 43.6 & 28.8 & 37.6 & 36.0 & 33.2 & 34.8 & 7.9 & 32.6 & 45.0 \\ \
ReasonIR with Rank-R1-32B & 38.8 & 59.5 & 55.1 & 37.9 & 52.7 & 30.0 & 39.3 & 45.1 & 32.1 & 17.1 & 10.7 & 40.4 & 45.6 \\ \
RaDeR with  QwenRerank & 39.2 & 58.0 & 59.2 & 33.0 & 49.4 & 31.8 & 39.0 & 36.4 & 33.5 & 33.3 & 10.8 & 34.2 & 51.6 \\ \
XRR2 & 40.3 & 63.1 & 55.4 & 38.5 & 52.9 & 37.1 & 38.2 & 44.6 & 21.9 & 35.0 & 15.7 & 34.4 & 46.2 \\ 
ReasonRank &40.8 &	62.7 & 55.5 & 36.7 & 54.6 & 35.7 & 38.0 & 44.8 & 29.5 & 25.6 & 14.4 & 42.0 & 50.1
\\ 
\textbf{DIVER(v1)} & 41.6 & 62.2 & 58.7 & 34.4 & 52.9 & 35.6 & 36.5 & 42.9 & \textbf{38.9} & 25.4 & 18.3 & 40.0 & 53.1 \\
BGE-Reasoner & 45.2 & 66.5 & \textbf{63.7} & 39.4 & 50.3 & 37.0 & 42.9 & 43.7 & 35.1 & 44.3 & 17.2 & 44.2 & \textbf{58.5}
 \\
\textbf{DIVER(v2)} & 45.8 & \textbf{68.0} & 62.5 & 42.0 & \textbf{58.2} & \textbf{41.5} & 44.3 & 49.2 & 34.8 & 32.9 & \textbf{19.1} & 44.3 & 52.6 \\
\textbf{DIVER(v3)} & \textbf{46.8} & 66.0 & \textbf{63.7} & \textbf{42.4} & 55.0 & 40.6 & \textbf{44.7} & \textbf{50.4} & 32.5 & \textbf{47.3} & 17.2 & \textbf{46.4} & 55.6 \\
\bottomrule
\end{tabular}
}
\label{tab:query_retriever_comparison}
\end{table}

\subsection{Ablation Studies}
This section investigates the contributions of three key components in our framework: the DIVER-Retriever for retrieval, DIVER-QExpand for query expansion, and DIVER-DChunk for document cleaning and chunking. 

\subsubsection{Comparison with Different Retrievers} \label{sec:perf_different_retriever}
We compare our DIVER-Retriever with a range of retrieval baselines: (1) the sparse retriever \textbf{BM25}~\citep{2009_bm25}; (2) open-source dense retrievers including \textbf{SBERT} (all-mpnet-base-v2, \cite{19emnlp_sBERT}), \textbf{gte-Qwen1.5-7B}~\citep{gte_qwen1.5}, and \textbf{Qwen3-4B}~(Qwen3-Embedding-4B, \cite{qwen3emb}); (3) proprietary models from \textbf{OpenAI}\footnote{\url{ https://openai.com/index/new-embedding-models-and-api-updates/.}} (text-embedding-3-large) and \textbf{Google} (text-embedding-preview0409, \cite{24google_gecko}); and (4) reasoning-aware retrievers including \textbf{ReasonIR-8B}~\citep{25_ReasonIR}, \textbf{RaDeR-7B}~\citep{25_RaDeR}, and \textbf{Seed1.5-Embedding}\footnote{\url{https://seed1-5-embedding.github.io/}}. 

As shown in \cref{tab:retriever_performance}, DIVER-Retriever achieves SOTA performance on the BRIGHT benchmark. It consistently outperforms other retrievers across original and expanded queries, with nDCG@10 scores of 28.9, 32.1, and 33.9, respectively. Notably, our 4B model significantly surpasses reasoning-aware retrievers such as ReasonIR-8B and RaDeR-7B, and even exceeds the commercial model SeedEmbedding-1.5 by 1.7 nDCG@10 on original queries. These results highlight DIVER-Retriever’s effectiveness in handling reasoning-intensive tasks. 
Furthermore, the performance of DIVER-Retriever can be further enhanced by integrating it with a sparse retrieval method. By interpolating its relevance scores with those from BM25 at a ratio of 0.5, the nDCG@10 score improves substantially to \textbf{37.2}. This hybrid scoring strategy highlights the complementary strengths of dense and sparse retrieval methods, offering a robust solution for capturing both deep reasoning and surface-level relevance in real-world retrieval scenarios.

\begin{table}[tbp]
\centering
\caption{nDCG@10 performance of various retrievers on the BRIGHT benchmark, evaluated using original queries, GPT-4 CoT reasoning or DIVER-QExpand queries. ``+BM25 (Hybrid)" denotes a variant that interpolates the similarity scores of the retriever and BM25 with an equal weight of 0.5. }

\resizebox{\textwidth}{!}{
\begin{tabular}{l|c|ccccccc|cc|ccc}
\toprule &
& \multicolumn{7}{c|}{\textbf{StackExchange}} 
& \multicolumn{2}{c|}{\textbf{Coding}} 
& \multicolumn{3}{c}{\textbf{Theorem-based}} \\
\textbf{Method} & \textbf{Avg.} 
& \textbf{Bio.} & \textbf{Earth.} & \textbf{Econ.} & \textbf{Psy.} & \textbf{Rob.} & \textbf{Stack.} & \textbf{Sus.}
& \textbf{Leet.} & \textbf{Pony} 
& \textbf{AoPS} & \textbf{TheoQ.} & \textbf{TheoT.} \\
\midrule
\multicolumn{14}{c}{Evaluate Retriever with Original Query} \\
\midrule
BM25 & 14.5 & 18.9 & 27.2 & 14.9 & 12.5 & 13.6 & 18.4 & 15.0 & 24.4 & 7.9 & 6.2 & 10.4 & 4.9 \\ 
SBERT & 14.9 & 15.1 & 20.4 & 16.6 & 22.7 & 8.2 & 11.0  & 15.3 & 26.4 & 7.0 & 5.3 & 20.0 & 10.8 \\ 
gte-Qwen1.5-7B & 22.5 & 30.6 & 36.4 & 17.8 & 24.6 & 13.2 & 22.2 & 14.8 & 25.5 & 9.9 & 14.4 & 27.8 & 32.9 \\ 
Qwen3-4B & 5.6 & 3.5 & 8.0 & 2.3 & 2.0 & 1.6 & 1.0 & 4.4 & 2.1 & 0.1 & 4.9 & 18.0 & 19.2 \\
OpenAI & 17.9 & 23.3 & 26.7 & 19.5 & 27.6 & 12.8 & 14.3 & 20.5 & 23.6 & 2.4 & 8.5 & 23.5 & 11.7 \\ 
Google & 20.0 & 22.7 & 34.8 & 19.6 & 27.8 & 15.7 & 20.1 & 17.1 & 29.6 & 3.6 & 9.3 & 23.8 & 15.9 \\ 
ReasonIR-8B & 24.4 & 26.2 & 31.4 & 23.3 & 30.0 & 18.0 & 23.9 & 20.5 & 35.0 & 10.5 & \textbf{14.7} & 31.9 & 27.2 \\ 
RaDeR-7B & 25.5 & 34.6 & 38.9 & 22.1 & 33.0 & 14.8 & 22.5 & 23.7 & 37.3 & 5.0 & 10.2 & 28.4 & 35.1 \\ 
Seed1.5-Embedding & 27.2 & 34.8 & 46.9 & 23.4 & 31.6 & 19.1 & 25.4 & 21.0 & \textbf{43.2} & 4.9 & 12.2 & 33.3 & 30.5 \\ 
DIVER-Retriever(v1) & 28.9 & 41.8 & 43.7 & 21.7 & 35.3 & 21.0 & 21.2 & 25.1 & 37.6 & 13.2 & 10.7 & \textbf{38.4} & 37.3 \\ 
DIVER-Retriever(v2) & \textbf{31.9} & \textbf{49.0} & \textbf{49.7} & \textbf{32.0} & \textbf{41.9} & \textbf{30.2} & \textbf{28.4} & \textbf{33.7} & 13.3 & \textbf{16.2} & 8.6 & 37.5 & \textbf{42.3} \\
\midrule
\multicolumn{14}{c}{Evaluate Retriever with GPT-4 REASON-query} \\
\midrule
BM25 & 27.0 & \textbf{53.6} & \textbf{54.1} & 24.3 & 38.7 & 18.9 & 27.7 & 26.3 & 19.3 & 17.6 & 3.9 & 19.2 & 20.8 \\ 
SBERT & 17.8 & 18.5 & 26.3 & 17.5 & 27.2 & 8.8 & 11.8 & 17.5 & 24.3 & 10.3 & 5.0 & 22.3 & 23.5 \\ 
gte-Qwen1.5-7B & 24.8 & 35.5 & 43.1 & 24.3 & 34.3 & 15.4 & 22.9 & 23.9 & 25.4 & 5.2 & 4.6 & 28.7 & 34.6 \\ 
Qwen3-4B & 5.5 & 1.3 & 17.3 & 2.5 & 6.2 & 1.0 & 4.8 & 4.5 & 3.0 & 5.9 & 0.0 & 7.2 & 12.5 \\ 
OpenAI & 23.3 & 35.2 & 40.1 & 25.1 & 38.0 & 13.6 & 18.2 & 24.2 & 24.5 & 6.5 & 7.7 & 22.9 & 23.8 \\ 
Google & 26.2 & 36.4 & 45.6 & 25.6 & 38.2 & 18.7 & \textbf{29.5} & 17.9 & 31.1 & 3.7 & 10.0 & 27.8 & 30.4 \\ 
ReasonIR-8B & 29.9 & 43.6 & 42.9 & \textbf{32.7} & 38.8 & 20.9 & 25.8 & \textbf{27.5} & 31.5 & \textbf{19.6} & 7.4 & 33.1 & 35.7 \\ 
RaDeR-7B & 29.2 & 36.1 & 42.9 & 25.2 & 37.9 & 16.6 & 27.4 & 25.0 & \textbf{34.8} & 11.9 & \textbf{12.0} & 37.7 & \textbf{43.4} \\ 
DIVER-Retriever(v1) & \textbf{32.1} & 51.9 & 53.5 & 29.5 & \textbf{41.2} & \textbf{21.4} & 27.5 & 26.1 & 33.5 & 11.7 & 9.5 & \textbf{39.3} & 39.7 \\
\midrule
\multicolumn{14}{c}{Evaluate retriever with DIVER-QExpand query} \\
\midrule
ReasonIR-8B & 32.6 & 49.4 & 44.7 & 32.4 & 44.0 & 26.6 & 31.8 & 29.0 & 32.3 & 12.8 & 9.1 & \textbf{40.7} & 38.4 \\ 
{ +BM25 (Hybrid)} & 35.7 & 56.8 & 53.5 & \textbf{33.0} & \textbf{48.5} & \textbf{29.4} & \textbf{34.2} & \textbf{32.0} & \textbf{35.2} & 16.8 & 12.9 & 39.3 & 36.8 \\
\textbf{DIVER-Retriever(v1)} & 33.9 & 54.5 & 52.7 & 28.8 & 44.9 & 25.1 & 27.4 & 29.5 & 34.5 & 10.0 & 14.5 & \textbf{40.7} & 44.7 \\
\textbf{ +BM25 (Hybrid)} & \textbf{37.2} & \textbf{60.0} & \textbf{55.9} & 31.8 & 47.9 & 27.1 & 33.9 & 31.9 & 35.1 & \textbf{23.1} & \textbf{16.8} & 36.9 & \textbf{46.6} \\
\bottomrule
\end{tabular}
}
\label{tab:retriever_performance}
\end{table}

\subsubsection{Comparison with Different Expanded Queries}
This section compares DIVER-QExpand with several representative zero-shot query expansion methods:
(1) \textbf{GPT-4 Reason-query}~\citep{25ICLR_BRIGHT}, which uses chain-of-thought reasoning steps from GPT-4 as queries, provided in BRIGHT;
(2) \textbf{XRR2}\footnote{\url{https://github.com/jataware/XRR2/tree/main}}, which applies carefully crafted prompts with GPT-4o for query expansion;
(3) \textbf{TongSearch-QR-7B}~\citep{25_tongsearch_qr}, a small-size language model trained by reinforcement learning~(RL) for reasoning-based query rewriting;
and (4) \textbf{ThinkQE-14B}~\citep{25_ThinkQE}, a thinking-based query expansion method that iteratively refines queries using feedback from retrieved documents.


\cref{tab:query_expansion_performance} shows that DIVER-QExpand achieves the highest overall performance on the BRIGHT benchmark, reaching an average nDCG@10 of \textbf{32.6} with the ReasonIR retriever, \textbf{1.8} points higher than the best baseline, ThinkQE. The gain comes from its feedback-based expansion strategy, where a stronger retriever provides more relevant context, enabling more effective query reformulation. This advantage is most evident with reasoning-aware retrievers, as DIVER-QExpand consistently outperforms competitive methods across most datasets when paired with ReasonIR. With BM25, it scores 29.5, slightly below ThinkQE, likely because ThinkQE’s longer expansions yield more keyword overlaps for lexical retrieval. Overall, DIVER-QExpand proves particularly effective in reasoning-intensive retrieval, where feedback during expansion offers clear advantage over other methods.
\begin{table}[tbp]
\centering
\caption{nDCG@10 performance of various query expansion methods on the BRIGHT benchmark, evaluated using BM25 and ReasonIR retrievers. }
\resizebox{\textwidth}{!}{
\begin{tabular}{l|c|ccccccc|cc|ccc}
\toprule &
& \multicolumn{7}{c|}{\textbf{StackExchange}} 
& \multicolumn{2}{c|}{\textbf{Coding}} 
& \multicolumn{3}{c}{\textbf{Theorem-based}} \\
\textbf{Method} & \textbf{Avg.} 
& \textbf{Bio.} & \textbf{Earth.} & \textbf{Econ.} & \textbf{Psy.} & \textbf{Rob.} & \textbf{Stack.} & \textbf{Sus.}
& \textbf{Leet.} & \textbf{Pony} 
& \textbf{AoPS} & \textbf{TheoQ.} & \textbf{TheoT.} \\
\midrule
\multicolumn{14}{c}{Using BM25 retriever} \\
\midrule
GPT-4 Reason-query & 26.5 & 53.6 & 53.6 & 24.3 & 38.6 & 18.8 & 22.7 & 25.9  & 19.3 & 17.7  & 3.9 & 18.9 & 20.2 \\ 
XRR2 & 24.5 & 57.7 & 52.5 & 22.1 & 36.0 & 17.0 & 23.9 & 25.3 & 16.3 & 11.5 & 2.4 & 15.8 & 14.2 \\ 
TongSearch-QR-7B & 27.9 & \textbf{57.9} & 50.9 & 21.9 & 37.0 & 21.3 & 27.0 & 25.6 & 23.6 & 14.4 & 7.0 & 26.1 & \textbf{22.0}  \\ 
ThinkQE & \textbf{30.0} & 55.9 & 52.3 & \textbf{26.5} & 39.0 & 22.9 & \textbf{27.9} & \textbf{30.9} & 25.2 & \textbf{20.9} & \textbf{10.3 }& \textbf{27.0} & 21.4 \\ 
DIVER-QExpand & 29.5 & 56.7 & \textbf{54.5} & 25.9 & \textbf{43.9} & \textbf{23.2} & 27.0 & 28.8 & \textbf{25.6} & 16.6 & 8.7 & 23.4 & 20.4 \\
\midrule
\multicolumn{14}{c}{Using ReasonIR retriever} \\
\midrule
GPT-4 Reason-query & 29.9 & 43.6 & 42.9 & \textbf{32.7} & 38.8 & 20.9 & 25.8 & 27.5 & 31.5 & 19.6 & 7.4 & 33.1 & 35.7 \\ 
XRR2 & 30.8 & 47.1 & \textbf{46.2} & 29.6 & 39.9 & 22.3 & 32.8 & 24.6 & 24.7 & \textbf{27.7} & 7.0 & 33.6 & 33.8 \\ 
TongSearch-QR-7B & 31.8 & 46.2 & 45.1 & 31.2 & 39.6 & 25.3 & 28.7 & 28.4 & 31.2 & 16.3 & \textbf{10.8} & 40.0  & \textbf{39.3} \\ 
ThinkQE & 30.8 & 44.9 & 45.2 & 31.9 & 40.9 & 24.6 & \textbf{33.5} & 28.9 & 23.5 & 9.3 & 8.5 & \textbf{41.1} & 37.4 \\ 
DIVER-QExpand & \textbf{32.6} & \textbf{49.4} & 44.7 & 32.4 & \textbf{44.0} & \textbf{26.6} & 31.8 & \textbf{29.0} & \textbf{32.3} & 12.8 & 9.1 & 40.7 & 38.4 \\
\bottomrule
\end{tabular}
}
\label{tab:query_expansion_performance}
\end{table}

\subsubsection{Effect of DIVER-DChunk} \label{sec:ablation_diver_dchunk}
We investigate the impact of DIVER-DChunk, the document cleaning and chunking strategy in our pipeline. In the main comparisons above, this component was not used to ensure fairness, as other baselines relied on the original BRIGHT document chunks without modification. However, the original corpus contains notable segmentation issues, particularly in seven StackExchange-derived subdomains, where structural artifacts from web scraping lead to inconsistent formatting, fragmented sentences
We applied DIVER-DChunk to these subdomains, using DIVER-QExpand queries. 

As shown in Table~\ref{tab:diver_dchunk_performance}, the improvement for DIVER-Retriever is consistent across most domains, with larger gains in Psychology (+2.3) and  Stack Overflow (+1.2). This suggests that cleaner and semantically aligned chunks provide richer context for dense retrieval models. The impact on BM25 is minimal, which is expected since it is less sensitive to chunk coherence. Our observation aligns with recent work~\citep{25_bright+} that also highlights the limitations of BRIGHT’s original chunking and proposes BRIGHT+, a multi-agent LLM-based cleaning and rechunking pipeline to build a higher-quality corpus.

\begin{table}[tbp]
\centering
\caption{Effect of DIVER-DChunk on retrieval performance across StackExchange domains. }
\begin{tabular}{l|c|ccccccc}
\toprule &
& \multicolumn{7}{c}{\textbf{StackExchange}} \\
\textbf{Method} & \textbf{Avg.} 
& \textbf{Bio.} & \textbf{Earth.} & \textbf{Econ.} & \textbf{Psy.} & \textbf{Rob.} & \textbf{Stack.} & \textbf{Sus.}\\
\midrule
bm25 & 37.1 & 56.7 & 54.5 & 25.9 & 43.9 & 23.2 & 27.0 & 28.8 \\ 
w/ DIVER-DChunk & 37.0 & 56.2 & 55.2 & 25.7 & 44.7 & 23.1 & 25.8 & 28.2 \\ 
\midrule
DIVER-Retriever & 37.5 & 54.5 & 52.7 & 28.8 & 44.9 & 25.1 & 27.4 & 29.5 \\
w/ DIVER-DChunk & 38.0 & 54.6 & 53.3 & 29.2 & 47.2 & 24.7 & 28.6 & 28.5 \\
\bottomrule
\end{tabular}

\label{tab:diver_dchunk_performance}
\end{table}


\section{Conclusion}
We present DIVER, a retrieval pipeline designed for reasoning-intensive tasks where query–document relevance hinges on implicit and abstract relationships.  On the BRIGHT benchmark, DIVER achieves an nDCG@10 of 45.8, surpassing the previous best score of 45.2 from BGE-Reasoner, and consistently outperforms strong baselines such as ReasonIR and RaDeR with both original and rewritten queries. These improvements largely stem from iterative query refinement and reasoning-oriented retrieval trained on high-quality hard contrastive document pairs.
Future work will focus on developing an end-to-end framework that integrates query expansion, retrieval, and reranking into a unified model, reducing system latency and computational overhead while maintaining high retrieval performance.

\clearpage
\appendix

\section{Appendix}
\label{sec:appendix}

This section details our strategy for constructing positive and negative samples when training embedding models. This method aims to enhance the model's discriminative power and generalization performance by leveraging the robust capabilities of large language models (LLMs) to generate high-quality positive samples and challenging hard negative samples.

\subsection{General sample construction}
To construct high-quality positive and negative samples for embedding model training, we first collect query–document pairs from a large-scale web search engine. Each query is paired with its retrieved document (denoted as Doc) to form an initial candidate set.

We then use GPT-4 to provide fine-grained relevance annotations for each query–document pair on a 0–10 scale, where higher scores indicate stronger semantic and contextual relevance. Specifically, pairs with annotation scores greater than 6 are selected as positive samples, reflecting high semantic alignment, while pairs with scores below 4 are selected as negative samples, indicating low or no semantic relevance. Samples with scores between 4 and 6 (inclusive) are excluded to ensure a clear distinction between the positive and negative classes, thereby improving the robustness of downstream contrastive learning.

This annotation process enables the embedding model to learn from both high-quality positive examples and challenging negative examples, facilitating better generalization and retrieval performance in real-world scenarios. The instruction given to the LLM is as follows:

\begin{promptbox}[prompt:general_sample_construct]{Prompt for General Sample Construction}
\label{prompt:general_sample_construct}
Your task is to judge how useful a piece of Doc is as a reference for answering a Query. The Query is the user's question; the Doc includes the web page's title and some retrieved snippets from the page.\\

Please follow the rules below strictly:

1. Pay attention to whether the time, place, subject, and object in the Query match those in the Doc; if they do not match, you must deduct points.

2. Pay special attention to whether proper nouns in the Query match those in the Doc;

3. Regarding the Doc:

3.1 Identify the main meaning of the Doc. If only a small part of the Doc is relevant while the majority discusses other topics, you must deduct points;

3.2 Assess the applicability of the Doc; if it is overly one-sided, you must deduct points.

4. If the Query is vague and its specific meaning cannot be determined, you should deduct points appropriately.

5. Your output must be in JSON format and contain 2 keys: one is score, and the other is reason. The score represents the number, and reason explains your scoring rationale. Do not output any other unrelated Doc.\\

I will now give you a Query and Doc. Please follow the rules above strictly and output the score and reason in JSON format. Do not output anything else.

Query: XXX

Doc: XXX

Response:
\end{promptbox}

\subsection{Hard sample generation}
\subsubsection{Positive Sample Generation}
Our approach to constructing positive samples focuses on capturing deep semantic relevance between queries and documents. We configure a pre-trained LLM, specifically Deepseek-R1, to simulate the role of a search engine. Given a specific query, the LLM is instructed to "generate" several highly relevant documents that a search engine would ideally retrieve for that query.

\textbf{Positive Sample Relevance Criteria.} When generating documents, the LLM's primary objective is to ensure that the output is semantically strongly related to the query, mimicking the behavior of a real search engine returning the most pertinent results. These documents may contain keywords directly associated with the query, but more crucially, they exhibit a high degree of conceptual or thematic consistency. The following box presents the instruction for generating positive samples.

\begin{promptbox}[prompt:pos_sample_construct]{Prompt for Positive Sample Generation}
\label{prompt:pos_sample_construct}
You are a simulated Google search engine. Your task is to return webpages that can answer the given Query. Please follow these guidelines: \\

1. Mimic the style of a typical webpage: each result must include both a title and content.

2. For multi-hop questions, you only need to generate the document for the final hop.  

Example: If the Query describes symptoms and asks for treatment, first infer the disease yourself, then produce a webpage that discusses treatment for that disease only—do not describe the symptoms again.

3. If the user’s intent is narrow and could reasonably be satisfied by a single webpage, generate just one document. If the Query contains multiple sub-questions that would normally require separate sources, provide multiple documents, ensuring each covers a distinct, non-overlapping topic.

4. Output format:  

Document 1:\{"title":"xxx","content":"xxx"\}

…  

Document n:\{"title":"xxx","content":"xxx"\}

5. Generate no more than three documents in total. The content of each document must be 400–800 words, self-contained, and coherent.

6. Remember, you are simulating real Google search results. Your webpages should not analyze or directly answer the Query; instead, they should present relevant information in a conversational, everyday style, similar to popular health sites like Dingxiangyuan—but without describing specific patient cases. \\

Now I will give you a Query; please generate webpage content in the required format.

Query: XXX

Response: 
\end{promptbox}

Each document generated by the LLM is treated as a positive sample for its corresponding query. This method ensures that positive samples not only share lexical overlap with the query but also provide rich contextual and relevant information at a semantic level.

\subsubsection{Hard-Negative Sample Generation}
Constructing hard negatives is essential for training embedding models, especially when we want them to discern subtle semantic differences. We therefore impose strict constraints on the LLM when it produces such samples.

1. LLM-generated hard negatives  
   For each query, the LLM is instructed to create negative samples. Unlike positive samples, these must adhere to explicit non-relevance criteria.

2. Core constraint  
   The generated negatives may share only superficial lexical overlap with the query; they must exhibit no deep semantic relevance.

3. Detailed requirements  
   a) Shallow lexical overlap: A hard negative may reuse a handful of keywords or short phrases from the query. This surface-level similarity mimics “deceptive” real-world documents that look relevant but are not.  
   b) No deep semantic relation: Despite the overlapping words, the LLM must ensure the overall meaning, topic, or context is entirely unrelated to the query.  
      Example: If the query is “How to grow roses,” an acceptable hard negative might mention “rose,” yet discuss “the popularity of rose-colored lipstick” or “the history of the Wars of the Roses,” rather than gardening.

This strategy compels the embedding model to distinguish between lexical resemblance and true semantic relevance. By exposing the model to these “seemingly related but actually irrelevant” samples, it learns to capture the genuine intent behind a query and relies less on simple keyword matching, thereby boosting retrieval accuracy in complex scenarios.
The following box presents the instruction for generating negtive samples.

\begin{promptbox}[prompt:hardneg_sample_construct]{Prompt for Hard-Negative Sample Generation}
\label{prompt:hardneg_sample_construct}
You have been assigned a paragraph-generation task:  

You will receive incomplete data containing the following information: 

• “input”: a string consisting of a random prompt specified by the task.  

• “positive document”: a string that, according to the task, is relevant to the “input.” \\

Your job is to produce a JSON-formatted “hard negative document”: 

• The “hard negative document” is a difficult negative sample. While it shares some lexical overlap with the input, it does not help solve the input’s problem and is less relevant to the input than the “positive document.” \\

Please observe these guidelines:  

1. The value of “hard negative document” must be written in Chinese. 

2. The “hard negative document” should be a long passage (at least 300 Chinese characters) and should avoid excessive lexical overlap; otherwise, the task will be too simple.  

3. The “input,” “positive document,” and “hard negative document” must remain independent of one another. \\

Your output must always be a single JSON object—provide no explanations or additional text. Be creative! \\

Now, apply the instructions to the following data:  

'input': XXX

'positive document': XXX

Your response:
\end{promptbox}

\clearpage
\bibliography{diver}
\bibliographystyle{colm2024_conference}

\end{document}